\documentclass[aps,superscriptaddress]{revtex4}
\usepackage{amssymb,amsmath,epsfig}
\usepackage{subfigure}
\usepackage{wrapfig}
\usepackage{float}
\graphicspath{{fig/}}
\usepackage[colorlinks=true, pdfstartview=FitV, linkcolor=blue, citecolor=red, urlcolor=magenta, breaklinks=true]{hyperref}
\begin{document}
\title{Absorption and scattering of a black hole with a global monopole in $ f(R) $ gravity}

\author{M. A. Anacleto}\email{anacleto@df.ufcg.edu.br}
\affiliation{Departamento de F\'{\i}sica, Universidade Federal de Campina Grande
Caixa Postal 10071, 58429-900 Campina Grande, Para\'{\i}ba, Brazil}

\author{F. A. Brito}\email{fabrito@df.ufcg.edu.br}
\affiliation{Departamento de F\'{\i}sica, Universidade Federal de Campina Grande
Caixa Postal 10071, 58429-900 Campina Grande, Para\'{\i}ba, Brazil}
\affiliation{Departamento de F\'isica, Universidade Federal da Para\'iba, 
Caixa Postal 5008, 58051-970 Jo\~ao Pessoa, Para\'iba, Brazil}
 
\author{S. J. S. Ferreira}\email{stefanejudith@gmail.com}
\affiliation{Departamento de F\'{\i}sica, Universidade Federal de Campina Grande
Caixa Postal 10071, 58429-900 Campina Grande, Para\'{\i}ba, Brazil}

\author{E. Passos}\email{passos@df.ufcg.edu.br}
\affiliation{Departamento de F\'{\i}sica, Universidade Federal de Campina Grande
Caixa Postal 10071, 58429-900 Campina Grande, Para\'{\i}ba, Brazil}

\begin{abstract} 
In this paper we consider the solution of a  black hole with a  global monopole in $ f(R)$ gravity 
and  apply the partial wave approach to compute the differential scattering cross section and absorption cross section. 
We show that in the low-frequency limit and at small angles the contribution to the dominant term in the scattering/absorption cross section is modified by the presence of the global monopole and the gravity modification. In such limit, the absorption cross section shows to be proportional to the area of the event horizon.
\end{abstract}

\maketitle
\pretolerance10000

\section{Introduction}
Black holes are  fascinating objects that have remarkable characteristics and one of them is that they behave like thermodynamic systems possessing temperature and entropy.
Black holes are exact solutions of Einstein equations which are determined by mass ($ M $), electric charge ($ Q $) 
and angular momentum ($ J $)~\cite{Frolov,Townsend1997} and plays an important role in modern physics. 
In particular, black hole with a global monopole has been explored extensively by many authors in recent years~\cite{BezerradeMello:1996si,Yu2002,Paulo2009,Chen2008,Rahaman2005}
and the metric for this kind of black hole was determined by Barriola and Vilenkin~\cite{Barriola1989}. 
The global monopoles are topological defects that arise in gauge theories due to the spontaneous symmetry breaking 
of the original global $ O(3) $ symmetry to $U(1)$~\cite{Kibble1976,Vilenkin1988}.
It is a type of defect that could be formed during phase transitions in the evolution of the early Universe.
From the cosmological point of view, the so-called $ f(R) $ theory of gravity introduces the possibility to explain the accelerated-inflation problem without the need to consider dark matter or dark energy~\cite{Nojiri,Carrol2004,Fay2007,Bazeia:2007jj}. 
In~\cite{Carames2011} it has been investigated the classical motion of a massive test particle in the gravitational field of a 
{global monopole in  $f(R)$ gravity}.
The authors in~\cite{Graca:2015jea} have calculated, using the WKB approximation,  the quasinormal modes for a black hole with a global monopole in $f(R)$ theory of gravity.
The thermodynamics of the black holes with a global monopole in $ f(R) $ gravity was discussed in~\cite{Man2013,Lustosa:2015hwa} and was treated analytically in~\cite{Man2015} the case of trong gravitational lensing for a massive source with a global monopole in $f(R)$ {theory of gravity}.
The main objective of this work is to compute the scattering cross section due to a black hole with a global monopole 
in $ f(R) $ gravity theory. In~\cite{Hai:2013ara} was analayzed the absorption problem for a massless scalar field propagating in general static spherically-symmetric black holes with a global monopole. 
{The geometry related to topological defects such as global monopoles is normally associated with a solid deficit angle characterized by some coupling parameter. As a Schwarzschild black hole swallow a global monopole its geometry is also affected and inherits a solid deficit angle. As a consequence the event horizon is modified and so do all the properties that follows from this phenomenon. Then it is natural to investigate how deep the differential scattering/absorption cross section of the black-hole global-monopole system is affected. Previous studies have pointed that the global monopole tends to increase these quantities for sufficiently large global monopole coupling parameter. In our present study we extend previous analysis in Einstein gravity to $f(R)$ gravity. It is natural to investigate the interplay between the parameters from global monopole and  $f(R)$ gravity to uncover the ultimate physical consequences by analyzing the differential scattering/absorption cross section of the black-hole global-monopole system.}

The study to understand the processes of absorption and scattering in the vicinity of black holes is one of the most important issues in theoretical physics and also of great relevance for experimental research.
We can explore the dynamics of a black hole by trying to disturb it away from its stationary configuration. 
Thus, examining the interaction of fields with black holes is of great importance  to understand aspects about formation, stability, and gravitational wave emission.
For many years, several theoretical works have been done to investigate the black hole scattering~\cite{Futterman1988} (see also references therein). 
Since 1970, many works have shown that at the long wavelength limit  
($ GM\omega\ll 1 $)~\cite{Matzner1977,Westervelt1971,Peters1976,Sanchez1976,Logi1977,Doram2002,Dolan:2007ut,Crispino:2009ki}, the differential scattering cross section for small angles presents the following result: $ d\sigma/d\Omega\approx 16G^2M^2/\theta^4 $. 
In addition, the calculation to obtain the low energy absorption cross section has been studied extensively in the literature~\cite{Churilov1974,Gibbons1975,Page1976,Unruh1976}.
Thus, in this case the absorption cross section in the long-wavelength limit of a massless neutral scalar field is  equal to the area of the horizon, $\sigma=4\pi r^2_h=16\pi G^2 M^2$~\cite{Churilov1973}.
On the other hand, for fermion fields one has been shown by Unruh~\cite{Unruh1976} that the absorption cross section is $ 2\pi G^2M ^2 $ in the low-energy limit. The result is exactly $1/8$ of that for the scalar wave in the low-energy limit.
An extension of the calculation of the absorption cross section for acoustic waves was performed in~\cite{Crispino:2007zz},~\cite{Dolan:2009zza} and~\cite{Oliveira:2010zzb}. 
The partial wave approach has also been extended to investigate the scattering by an acoustic black hole in $(2 + 1)$ dimensions~\cite{Dolan,ABP2012-1,Anacleto:2015mta} and also due to a non-commutative BTZ black hole~\cite{Brito2015}.
Also some studies have been carried out on the processes of absorption and scattering of massive fields by black holes~\cite{Jung2004,Doran2005,Dolanprd2006,Castineiras2007,Benone:2014qaa}. For computations of scattering by spherically symmetric $d$-dimensional black holes in string theory see, e.g., \cite{Moura:2011rr}.

In this paper, inspired by all of these previous works and adopting the technique developed by the authors in~\cite{Dolan,ABP2012-1,Anacleto:2015mta,Brito2015,Marinho:2016ixt}, we shall focus on the computation of the scattering and absorption cross section for a monochromatic planar wave of neutral massless scalar field impinging upon a  black hole with a {global monopole in $f(R)$ gravity}. 
In this scenario there are four parameters: the mass $ M $ of the black hole, the frequency $ \omega $ of the field, the monopole parameter $ \eta $ and $ \psi_0 $ associated with the corrections from the $f(R)$ gravity. 
Thus, we have three dimensionless parameters: $ GM\omega $, $ 8\pi G\eta^2\approx 10^{-5} $ and $ a=\omega/\psi_0 $.
In our analyzes, we will consider only the long-wavelength regime, in which $ GM\omega\ll 1 $. 
Dolan et al.~\cite{Dolan}, studied the analogous Aharonov-Bohm effect considering the scattering of planar waves by a
draining bathtub vortex. They implemented an approximation formula  to calculate the phase shift $\delta_{l} \approx (m-\tilde{m})$ analytically, 
{where $ \tilde {m} $ was defined by considering only the contributions of $ m $ and $ \omega $ (frequency) appearing 
in the $ 1/r ^ 2 $ term modified after the power series expansion of $ 1/r $}. 
In an analogous way we introduce the following approximation: $ \delta_{l}\approx (l-\ell) $, 
{where $ \ell $ is defined by considering only the contributions of $ l $ and $ \omega $ appearing 
in the $ 1/r ^ 2 $ term modified after the power series expansion of $ 1/r $}.
Then, we have verified that the presence of the parameters $ \eta $ and $ \psi_0 $ modify the dominant term of the differential scattering cross section in the low-frequency limit at small angles and also the absorption cross section. 
We initially analyzed the example of the black hole with a global monopole and showed that the contribution to the dominant term of the differential cross section is increased due to the monopole effect as well as to the absorption.
{On the other hand, considering the case of a black hole with a global monopole in $ f(R) $ theory, we find that  in the low-frequency limit the contribution to the dominant term  of the differential scattering cross section  and for the absorption cross section  is also increased due to the effect of the $ f(R)$ theory.}
Here we adopt the natural units $ \hbar=c=k_B=1$.

\section{Scattering/Absorption Cross Section}
In this section we are interested in determining the differential scattering cross section for a black hole with a global monopole in $f (R) $ gravity by the partial wave method  in the low frequency regime. For this purpose we will follow the procedure adopted in previous works to calculate the phase shift. 

\subsection{The global monopole in Einstein gravity}
Initially, we will consider a spherically symmetric line element of a black hole with a global monopole that is given by
\begin{eqnarray}
\label{metresf}
ds^2=A(r)dt^2-\frac{dr^2}{A(r)}-r^2d\Omega^2,
\end{eqnarray}
where
\begin{eqnarray}
\label{metmonopole}
A(r)=1-8\pi G\eta^2 -\dfrac{2GM}{r}.
\end{eqnarray}
Here, $ G $ is the Newton constant, 
$ \eta $ is the monopole parameter of the order $ 10^{16}$GeV  and 
so $ 8\pi G\eta^2\approx 10^{-5} $~\cite{Vilenkin1985,Barriola1989}.
The event horizon radius is obtained by $ A(r)=0 $, i.e.
\begin{eqnarray}
\label{rheta}
r_{\eta}=\frac{2GM}{(1-8\pi G\eta^2)}=\frac{r_s}{(1-8\pi G\eta^2)},
\end{eqnarray}
{where $r_s=2GM$ is the event horizon of the Schwarzschild black hole.}

The Hawking temperature of the black hole is
\begin{eqnarray}
T_H=\frac{1}{4\pi}\left(\frac{1-8\pi G\eta^2}{r_s}\right).
\end{eqnarray}
For $ \eta=0 $ the Hawking temperature of the Schwarzschild black hole is recovered.

The next step is to consider the Klein-Gordon  wave equation for a massless scalar field in the background (\ref{metresf})
\begin{eqnarray}
\dfrac{1}{\sqrt{-g}}\partial_{\mu}\Big(\sqrt{-g}g^{\mu\nu}\partial_{\nu}\Phi\Big)=0 .
\end{eqnarray}
Now we can make a separation of variables into the equation above as follows
\begin{eqnarray}
\Phi_{\omega l m}({\bf r},t)=\frac{R_{\omega l}(r)}{r}Y_{lm}(\theta,\phi)e^{-i\omega t},
\end{eqnarray}
where $ \omega $ is the frequency and $Y_{lm}(\theta,\phi)  $ are the spherical harmonics.

In this case, the equation for $ R_{\omega l}(r) $ can be written as 
\begin{eqnarray}
\label{eqrad}
A(r)\dfrac{d}{dr}\left(A(r)\dfrac{dR_{\omega l}(r)}{dr} \right) +\left[ \omega^2 -V_{eff} \right]R_{\omega l}(r)=0,
\end{eqnarray}
and 
\begin{eqnarray}
V_{eff}=\frac{A(r)}{r}\frac{dA(r)}{dr}+\frac{A(r)l(l+1)}{r^2}, 
\end{eqnarray}
is the effective potential.
At this point, we  consider a new radial function, $ \psi(r)=A^{1/2}(r)R(r) $, so we have
\begin{eqnarray}
\label{eqradpsi}
\dfrac{d^2\psi(r)}{dr^2}+U(r) \psi(r) = 0,
\end{eqnarray}
where
\begin{eqnarray}
\label{poteff}
U(r)=\dfrac{[A'(r)]^2}{4 A^2(r)} - \dfrac{A''(r)}{2A(r)} + \dfrac{\omega^2}{A^2(r)} - \dfrac{V_{eff}}{A^2(r)},
\end{eqnarray}
and 
\begin{eqnarray}
&&A'(r)=\frac{dA(r)}{dr} =\dfrac{2GM}{r^{2}}  , \quad \quad A''(r)=\frac{d^2A(r)}{dr^2}= - \dfrac{4GM}{r^{3}} .
\end{eqnarray}
Now performing a power series in $1/r$ the Eq. (\ref{eqradpsi}) becomes
\begin{eqnarray}
\frac{d^2\psi(r)}{dr^2}+\left[\tilde{\omega}^2+{\cal V}(r)+ {\cal U}(r)\right] \psi(r) = 0,
\end{eqnarray}
where now we have 
\begin{eqnarray}
\label{pot1}
{\cal V}(r)= \frac{4GM\tilde{\omega}^2}{(1-8\pi G\eta^2)r}+\frac{12\ell^2}{r^2},
\end{eqnarray}
and 
\begin{eqnarray}
\label{pot2}
{\cal U}(r)&=&\frac{32 G^3 M^3\tilde{\omega}^3-2 ( l^2+l)G M\tilde{\omega}(1-8\pi G\eta^2)- 16\pi G\eta^2GM\tilde{\omega}}
{\tilde{\omega}(1-8\pi G\eta^2)^3r^3}
\nonumber\\
&+&\dfrac{1}{\tilde{\omega}^2(1-8\pi G\eta^2)^4r^4}\Big[80 G^4 M^4\tilde{\omega}^4+G^2 M^2\omega^2-4(l^2+l)G^2 M^2\omega^2(1-8\pi G\eta^2)
\nonumber\\
&-&(1-8\pi G\eta^2)\left[8-5(1-8\pi G\eta^2) \right]G^2 M^2\tilde{\omega}^2\Big]+\cdots,
\end{eqnarray}
{with $ \tilde{\omega}=\omega/(1-8\pi G\eta^2) $ and we define }
\begin{eqnarray}
\label{ell}
\ell^2\equiv-\frac{(l^2+l)}{12(1-8\pi G\eta^2)}+\frac{G^2M^2\tilde{\omega}^2}{(1-8\pi G\eta^2)^2}.
\end{eqnarray}
{Here $ \ell^2 $ was defined as the change of the coefficient of $1 / r ^ 2$ (containing only the contributions involving the quantities of $ l $ and $ \omega $) that arises after the realization of the power series in $ 1/r $ in Eq. (\ref{eqradpsi}).}
Notice that when $ r \rightarrow \infty $ the potential $ V(r)={\cal V }(r)+{\cal U}(r) \rightarrow 0 $ and the asymptotic behavior is satisfied.
Thus, knowing the phase shifts the scattering amplitude can be obtained  and which has the following partial-wave representation
\begin{eqnarray}
\label{ampl}
f(\theta)=\frac{1}{2i\omega}\sum_{l=0}^{\infty}(2l+1)\left(e^{2i\delta_l} -1 \right)P_{l}\cos\theta,
\end{eqnarray}
and the differential scattering cross section can be computed by the formula
\begin{eqnarray}
\dfrac{d\sigma}{d\theta}=\big|f(\theta) \big|^2.
\end{eqnarray}
The phase shift $ \delta_{l} $ can be obtained applying the folowing approximation formula
\begin{eqnarray}
\label{formapprox}
\delta_l\approx\frac{1}{2}(l-\ell)=\frac{1}{2}\left(l-\sqrt{-\frac{(l^2+l)}{12(1-8\pi G\eta^2)}+\frac{G^2M^2\tilde{\omega}^2}{(1-8\pi G\eta^2)^2}}   \right).
\end{eqnarray}
In the limit $ l\rightarrow 0 $ we obtain
\begin{eqnarray}
\label{phase2}
\delta_l=-\frac{GM\tilde{\omega}}{2(1-8\pi G\eta^2)}+{\cal O}(l)=-\frac{GM\omega}{2(1-8\pi G\eta^2)^2}+{\cal O}(l).
\end{eqnarray}
Note that in the limit $ l\rightarrow 0 $ the phase shifts tend to non-zero term, 
which naturally leads to a correct result for the differential cross section at the small angles limit.
Another way of obtaining the same phase shift is through the Born approximation formula
\begin{eqnarray}
\delta_l\approx\frac{\omega}{2}\int_0^{\infty}r^2J_l^2(\omega r) {\cal U}(r)dr,
\end{eqnarray}
where  $J_l(x)$ are the spherical Bessel functions of the first
kind and $ {\cal U}(r) $ is the effective potential of Eq. (\ref{pot2}).
After performing the integration we take the limits of $ \omega \rightarrow 0 $ and $ l\rightarrow 0 $. 
So the result is the same as Eq. (\ref{phase2}).

The Eq. (\ref{ampl}) is poorly convergent,  so it is very difficult to perform the sum of the series directly. This is due to the fact that an infinite number  of Legendre polynomials are required to obtain divergences in $ \theta=0 $. 
In~\cite{Yennie1954}, it has been found by the authors a way to around this problem. 
It has been proposed by them a reduced series which is less divergent in $ \theta=0 $, i.e.
\begin{eqnarray}
(1-\cos\theta)^{m}f(\theta)=\sum_{l=0}a_l^{m}P_{l}\cos\theta,
\end{eqnarray}
and so it is expected that the reduced series can converge more quickly.

Therefore, to determine the differential scattering cross section, we will use the following equation~\cite{Yennie1954,Cotaescu:2014jca}
\begin{eqnarray}
\label{espalh}
\dfrac{d\sigma}{d\theta}=\Big| \frac{1}{2i\tilde{\omega}}\sum_{l=0}^{1}(2l+1)\left(e^{2i\delta_l} -1 \right)
\frac{P_{l}\cos\theta}{1-\cos\theta}\Big|^2.
\end{eqnarray}
However,  considering few values of $ l $ ($l=0,1$) is sufficient to obtain the result satisfactorily.
Hence the differential scattering cross section is in this case given by
\begin{eqnarray}
\frac{d\sigma}{d\theta}\Big |^{\mathrm{l f}}_{\omega\rightarrow 0}=\frac{16G^2M^2}{\left( 1-8\pi G\eta^2\right)^2\theta^4}+\cdots 
=\frac{16G^2M^2}{\theta^4}\Big[ 1+16\pi G\eta^2+{\cal O}(G\eta^2)^2\Big] +\cdots .
\end{eqnarray}
The dominant term is modified by monopole parameter $ \eta $. 
Thus, we verified that the differential cross section is increased by the monopole effect.
As $ \eta=0 $ we obtain the result for the Schwarzschild black hole case.

{Now we will determine the absorption cross section for a black hole with a global monopole in the low-frequency  limit.}
As is well known in quantum mechanics, the total absorption cross section can be computed by means of the following relation
\begin{eqnarray}
\sigma_{abs}
=\frac{\pi}{\omega^2}\sum_{l=0}^{\infty}(2l+1)\Big(\big|1-e^{2i\delta_l}\big|^2\Big).
\end{eqnarray}
For the phase shift $ \delta_l $ of the Eq. (\ref{phase2}), we obtain in the limit $ \omega\rightarrow 0  $:
\begin{eqnarray}
\label{abs1}
\sigma_{abs}^{\mathrm{l f}}
&=&\frac{\pi}{\tilde{\omega}^2}\sum_{l=0}^{3}(2l+1)\Big(\big|1-e^{2i\delta_l}\big|^2\Big),
\nonumber\\
&=&\frac{16\pi G^2M^2}{\left( 1-8\pi G\eta^2\right)^2}=\frac{{\cal A}_{Sch}}{\left( 1-8\pi G\eta^2\right)^2},
\end{eqnarray}
where $ {\cal A}_{Sch} =4\pi r^2_s$ is the area of the event horizon of the Schwarzschild black hole.
So for a few values of $ l $ ($ l=0,1,2,3 $) the result is successfully obtained.
Here we note that the absorption is increased due to the contribution of the monopole.
{In~\cite{Hai:2013ara} the absorption cross section of a massless scalar wave due to a black hole with a global monopole has been computed. The authors have shown that the effect of the parameter $ \eta $ makes the black hole absorption stronger. 
Our result is in accordance with the one obtained in~\cite{Hai:2013ara}. } 
{Furthermore, our results for absorption show concordance} with the universality property of the absorption cross section which is always proportional to the area of the event horizon at low-frequency limit~\cite{Das:1996we}.
{In addition, in Fig. \ref{mla}  we show the graph for the mode $ l=0 $ of the absorption cross section that was obtained by numerically solving the radial equation (\ref{eqrad}) for arbitrary frequencies. }

\subsection{The global monopole in $f(R)$ gravity}
We will now compute the differential scattering cross section of a black hole with a global monopole in the 
$f(R)$ gravity. The spherical symmetric line element is given as follow~\cite{Carames2011, Lustosa:2015hwa, Chen:2016ftz}
\begin{eqnarray}
\label{metr}
ds^2=B(r)dt^2-\frac{dr^2}{B(r)}-r^2d\Omega^2,
\end{eqnarray}
where 
\begin{eqnarray}
\label{mfr}
B(r)=A(r)-\psi_0 r, \quad\quad A(r)=1-8\pi G\eta^2 -\dfrac{2GM}{r}.
\end{eqnarray}
 The term $ \psi_0 r $ corresponds to the extension of the standard general relativity.
For metric (\ref{metr}), when $ B(r)=0 $ {we obtain the following {\it internal} and {\it external} event horizons, respectively}
\begin{eqnarray}
\label{rh}
r_{-}=\frac{1-8\pi G\eta^2-\sqrt{(1-8\pi G\eta^2)^2-8GM\psi_0}}{2\psi_0},
\end{eqnarray}
and 
\begin{eqnarray}
\label{rc}
r_{+}=\frac{1-8\pi G\eta^2+\sqrt{(1-8\pi G\eta^2)^2-8GM\psi_0}}{2\psi_0}.
\end{eqnarray}
Adding $r_{-}$ and $r_{+}$ we also find the following relationship between horizons 
\begin{eqnarray}
\label{add-horizons}
 r_{+}+r_{-}=\frac{1}{\psi_0}\left(1-8\pi G\eta^2\right).
\end{eqnarray}
{Notice that the horizon $r_{+} $ exists only if $ \psi_0 $ is nonzero.}
Considering that $\psi_0$ is small and expanding the square root term in Eq.~(\ref{rh}) we obtain
\begin{eqnarray}
\label{rhschw}
r_{\eta}=\frac{2GM}{1-8\pi G\eta^2}+4G^2M^2\psi_0+\cdots=\frac{r_s}{1-8\pi G\eta^2}+r_s^2\psi_0+\cdots,
\end{eqnarray}
which for $ \psi_0=0 $ is exactly the result previously obtained in Eq. (\ref{rheta})  and when $ \eta=\psi_0=0 $ we have $r_{\eta}=r_s$,
that is the event horizon of the Schwarzschild black hole.  

{For Eq. (\ref{rc}), considering $ \psi_0 $ very small, we find  }
\begin{eqnarray}
\label{rcpsi}
r_{\psi_0}&=&\frac{1}{\psi_0}\left(1-8\pi G\eta^2\right)-\frac{r_s}{(1-8\pi G\eta^2)}-r_s^2\psi_0+\cdots,
\nonumber\\
&=&\frac{1}{\psi_0}\left(1-8\pi G\eta^2\right)-r_{\eta}+\cdots,\quad \Rightarrow \quad 
r_{\psi}+r_{\eta}=\frac{1}{\psi_0}\left(1-8\pi G\eta^2\right).
\end{eqnarray}
Observe that, at the limit of $ \psi_0\rightarrow 0 $, the effect of the theory $ f(R) $ has a dominant contribution only in the 
{external horizon $ r_{+} $.}
The Hawking temperature associated with the black hole of Eq.~(\ref{metr}) is
\begin{eqnarray}\label{ThRC}
T_H=\frac{1}{4\pi}\left(\frac{1-8\pi G\eta^2}{r_h} -2\psi_0 \right)
=\frac{1}{4\pi}(1-8\pi G\eta^2)\frac{r_{+}-r_{-}}{r_{-}(r_{+}+r_{-})},
\end{eqnarray}
{where in the last step of Eq.~(\ref{ThRC}) we have used Eq.~(\ref{add-horizons}).}
If $ \eta=0 $ and $ \psi_0=0 $ (in the absence of monopoles and $ f(R)$ corrections) the Hawking temperature will be reduced to that of the Schwarzschild case, as expected. 
However, we also note in (\ref{ThRC}) that for $T_H\geq0$, it is necessary that $r_{+}\geq r_{-}$, when $\psi_0$ is turned on. {This implies that the external horizon, $r_{+} $, should be larger than or equal the internal horizon $ r_{-} $. 
The temperature is zero when one saturates the lower bound, i.e., when $r_{+}=r_{-}$. This is an effect similar to what happens with Reissner-Nordstr\"om black holes}

Following the same steps applied to the previous case, Eq. (\ref{eqradpsi}) can be now written as
\begin{eqnarray}
\label{eqlpsi}
\frac{d^2\psi(r)}{dr^2}+\left[\frac{(\ell^2 +1/4)}{r^2}+ U(r)\right] \psi(r) = 0,
\end{eqnarray}
being
\begin{eqnarray}
U(r)&=&\frac{a(1 - 8 \pi G \eta^2)/2 + a (l^2+l + 1)+2(1 - 8 \pi G \eta^2) a^3}{\omega r^3} 
 + \frac{1}{\omega^2r^4} \Big[ -4 G M\omega (a^3+a) +(3 - 48 \pi G \eta^2)a^4  
\nonumber\\ 
 &+&  \frac{3a^2}{4} -8 \pi G \eta^2 a^2 + a^2 l (l + 1) (1 - 8 \pi G \eta^2)
 + 2a^2(1 - 8 \pi G \eta^2)\Big]+\cdots ,
\end{eqnarray}
where we have defined $ \ell^2\equiv a^2 $ and $ a\equiv\omega/\psi_0 $. 
{Analyzing the coefficient of $ 1 / r ^ 2 $ in Eq. (\ref{eqlpsi}), only the first term contains the frequency $ \omega $ and the second term is just a numerical factor. Thus, as already mentioned in Eq. (\ref{ell}), only the first term enters in the definition of $ \ell ^ 2 $.}
Note that the potential $V(r)={(\ell^2 +1/4)}/{r^2}+U(r) $ obeys the asymptotic limit $ V(r) \rightarrow 0$ as $ r \rightarrow \infty $.

Next following the same approximation used in the formula (\ref{formapprox}) the phase shift $ \delta_{l} $ in the limit $ l\rightarrow 0 $ reads
\begin{eqnarray}
\label{phase1}
\delta_l=-\frac{\omega}{2\psi_0}+ {\cal O}(l)
=-\frac{\bar{\omega}(r_{\psi}+r_{\eta})}{4} + {\cal O}(l),
\end{eqnarray} 
{where $ \bar{\omega} =\omega/2\left( 1-8\pi G\eta^2\right)$ and  we have used the result of Eq. (\ref{rcpsi}) to express the phase shift in terms of $ r_{\psi} $ and $ r_{\eta} $}. Also in this case the phase change tends to a non-zero constant term in the limit $ l\rightarrow 0 $.
Once again we can verify that the phase shift could have been obtained from the Born approximation formula
\begin{eqnarray}
\delta_l\approx\frac{\omega}{2}\int_0^{\infty}r^2J_l^2(\omega r)U(r)dr,
\end{eqnarray}
Thus,  in the low-frequency (long-wavelength) limit  and at the small angle $ \theta $, the differential scattering cross section is given by
\begin{eqnarray}
\dfrac{d\sigma}{d\theta}\Big |^{\mathrm{l f}}_{\omega\rightarrow 0}&=&\Big| \frac{1}{2i\bar{\omega}}\sum_{l=0}^{1}(2l+1)\left(e^{2i\delta_l} -1 \right)\frac{P_{l}\cos\theta}{1-\cos\theta}\Big|^2,
\nonumber\\
&=&\frac{4 \big(r_{\psi_0}+r_{\eta}\big)^2}{\theta^4}+\cdots.
\end{eqnarray}
{ In the limit $\psi_0\rightarrow 0  $  we have $ r_{\psi_0}\gg r_{\eta} $ and the differential scattering cross section becomes }
\begin{eqnarray}
\dfrac{d\sigma}{d\theta}\Big |^{\mathrm{l f}}_{\omega\rightarrow 0}
\approx\frac{4 r_{\psi}^2}{\theta^4}+\cdots
=\dfrac{4\left( 1-8\pi G\eta^2\right)^2}{\psi^2_0\theta^4}+\cdots.
\end{eqnarray}
We see that the presence of the parameters $ \psi_0 $ modifies the dominant term. 

{Now we will determine the absorption cross section for a black hole with a global monopole in $f (R)$  
gravity in the low-frequency  limit}.  So for the phase shift $ \delta_l $ (\ref{phase1}) and applying the limit $ \omega\rightarrow 0$ we find
\begin{eqnarray}
\label{absorpsi}
\sigma_{abs}^{\mathrm{l f}}&=&\frac{\pi}{\bar{\omega}^2}\sum_{l=0}^{3}(2l+1)\Big(\big|1-e^{2i\delta_l}\big|^2\Big),
\nonumber\\
&=& 4\pi\big( r_{\psi_0}+r_{\eta}\big)^2.
\end{eqnarray}
{In the limit $\psi_0\rightarrow 0  $  we have $ r_{\psi_0}\gg r_{\eta} $ and the absorption cross section becomes 
\begin{eqnarray}
\sigma_{abs}^{\mathrm{l f}}\approx \frac{4\pi\left( 1-8\pi G\eta^2\right)^2}{\psi_0^2}= 4\pi r_{\psi_0}^2
=A_{\psi_0},
\end{eqnarray}
which is dominated by the effect of the $ f(R)$ gravity. 
Thus, in this case the black hole with global monopole in $f(R)$ gravity absorbs more for large external horizon.
Finally, in the regime $ r_{\psi_0}=0 $ and $ \eta=0 $ from Eq.~(\ref{absorpsi}) we obtain the result for the case of the Schwarzschild black hole $\sigma_{abs}^{\mathrm{l f}}=4\pi r_{s}^2=A_{Sch}$}.

\begin{figure}[htbh]
\centering
\subfigure[]{\includegraphics[scale=0.93]{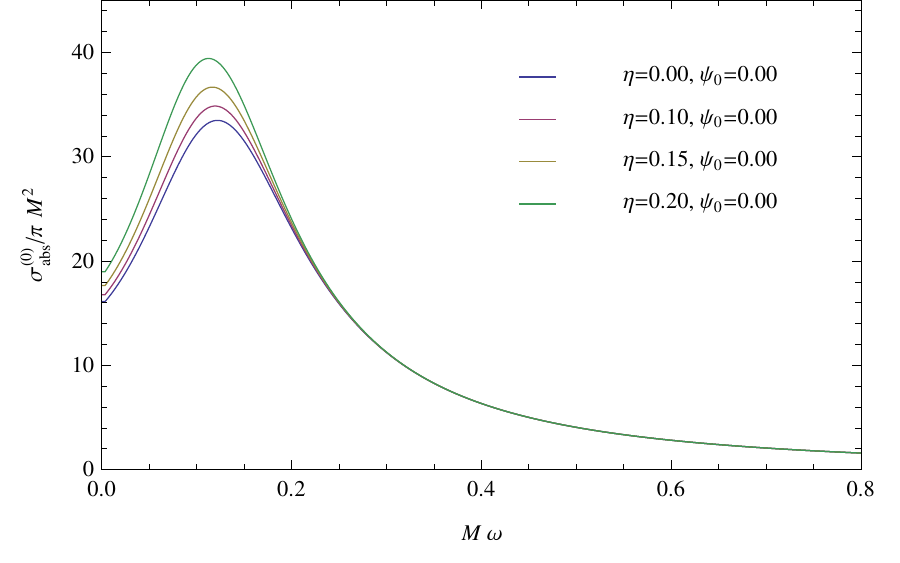}}\label{mla1}
\qquad
\subfigure[]{\includegraphics[scale=0.93]{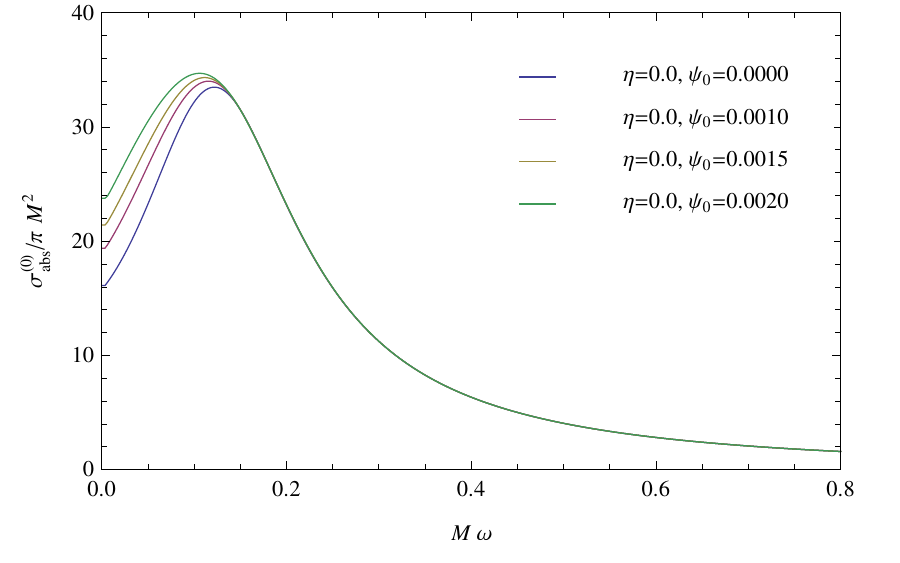}}\label{mla2}

\caption{Partial absorption cross section for the $l=0$ mode.}
\label{mla}
\end{figure}

{ It is worth mentioning that considering the metric $ B(r) $ in (\ref{mfr}), we note that in the limit when  $ r\rightarrow\infty $ we have $ B(r)\rightarrow\infty $. Thus, by analyzing the third term of the potential $ U(r) $ of Eq. (\ref{poteff}) we find that this term tends to zero when $ r $ goes to infinity. This is reflected in the absence of the frequency $ \omega^2 $ and the mass $ M $ in the term $1 / r ^ 2$ of Eq. (\ref{eqlpsi}) when we perform a series expansion of potential in $1 / r$. And as a consequence we do not have the presence of mass $ M $ in the result of absorption.  In order to avoid this and also to be in accordance with the numerical result we will consider the Eq. (\ref{eqradpsi}) after a variable change ($ 1/r \rightarrow 1+1/r $) written as follows:
\begin{eqnarray}
\label{Eqradfr}
\frac{d^2\psi(r)}{dr^2}+U(r) \psi(r) = 0,
\end{eqnarray}
where
\begin{eqnarray}
U(r)=\frac{\omega^2}{C^2(r)}+\frac{1}{C^2(r)}\left(\frac{[C'(r)]^2}{4} - \frac{C''(r)C(r)}{2} -\frac{C(r)C'(r)}{r}-\frac{C(r)l(l+1)}{r^2}\right),
\end{eqnarray}
being 
\begin{eqnarray}
C(r)=1-8\pi G\eta^2-\frac{2GM+\lambda}{r}-\frac{\psi_0}{r(1+r)}. 
\end{eqnarray} 
Notice that when $ r\rightarrow\infty $ the term $ \omega^2/C^2(r) $ tends to $ \omega^2 /(1-8\pi\eta^2)^2$.
Now writing the potential $ U(r) $ as a power series in $ 1/r $ we have
\begin{eqnarray}
\frac{d^2\psi(r)}{dr^2}+\left[\tilde{\omega}^2+ {\cal U}(r)\right] \psi(r) = 0,
\end{eqnarray}
where  $ \tilde{\omega}=\omega/(1-8\pi G\eta^2) $ and 
\begin{eqnarray}
{\cal U}(r)=\frac{\tilde{\omega}^2(4GM+\psi_0)}{(1 - 8 \pi G\eta^2) r}
+\frac{12\ell^2}{r^2}+\cdots,
\end{eqnarray}
here we have defined
\begin{eqnarray}
\label{ell-l}
\ell^2\equiv-\frac{(l^2+l)}{12(1-8\pi G\eta^2)}+\frac{\Big[G^2M^2+GM\psi_0+\psi_0(1-8\pi G\eta^2)/6+\psi_0^2/8)\Big]\tilde{\omega}^2}{(1-8\pi G\eta^2)^2}.
\end{eqnarray} 
Now we compute the phase shift through the approximation formula (\ref{formapprox}) in the limit $ l\rightarrow 0 $ and considering $ \psi_0 $ very small, which is given by 
\begin{eqnarray}
\label{phase3}
\delta_l=-\frac{GM\tilde{\omega}}{2(1-8\pi G\eta^2)}
\left[1+\frac{\psi_0\Big(G M+(1-8\pi G\eta^2)/6 \Big)}{2G^2 M^2}\right]+{\cal O}(l).
\end{eqnarray} 
Applying the formula (\ref{espalh}) we can obtain the following result for the differential scattering cross section
\begin{eqnarray}
\frac{d\sigma}{d\theta}\Big |^{\mathrm{l f}}_{\omega\rightarrow 0}=\frac{16G^2M^2}{\left( 1-8\pi G\eta^2\right)^2\theta^4}
\left[1+\frac{\psi_0\Big(G M+(1-8\pi G\eta^2)/6 \Big)}{2G^2 M^2}\right]^2+\cdots.
\end{eqnarray}
Note that the dominant term is modified by the parameters $ \eta $ and $ \psi_0 $.

In the low-frequency limit the absorption cross section reads
\begin{eqnarray}
\label{abs3}
\sigma_{abs}^{\mathrm{l f}}
=\frac{16\pi G^2M^2}{\left[ 1-8\pi G\eta^2\right)^2}
\left(1+\frac{\psi_0\Big(G M+(1-8\pi G\eta^2)/6 \Big)}{2 G^2 M^2}\right]^2.
\end{eqnarray}  
For $ \eta=0 $, we can verify that when we increase the value of $ \psi_0 $, 
the absorption has its value increased due to the effect of gravity $f (R)$. 
And when $ \psi_0=0 $, we recover the result of Eq. (\ref{abs1}). This can be best understood by looking at the graph of Fig. \ref{mla} 
for the mode $ l=0 $ which was obtained by numerically solving the radial equation (\ref{eqrad}) (with $ A(r) \rightarrow B(r) $) for arbitrary frequencies .}
 
 { At this point we present the numerical results of the partial absorption cross section as a function of arbitrary frequencies obtained through  the numerical procedure as described in \cite{Dolan:2012yc}. The graphs are shown below. }

In Fig.~\ref{mla}(a), we plot the partial absorption cross section for the $ l=0 $ mode with $ \eta=0.000, 0.100 $, $ 0.150 $ and 0.200. 
We can see by comparing the curves for different values of $\eta$ that the absorption is increased due to the contribution of the monopole. 
Moreover, when $ M\omega\rightarrow 0 $ the abostion tends to a nonzero value and when $ M\omega $ increases it tends to zero.
For $ \eta =0$ the graph shows the result of the partial absorption for the Schwarzschild black hole. 
Thus for non-zero values of $ \eta $, the partial absorption for the black hole with global monopole is increased in relation to the Schwarzschild black hole. Our result is in agreement with the one obtained in~\cite{Hai:2013ara}, for instance.

The effect of  $f (R)$ gravity for the partial absorption cross section for the $ l=0 $ mode can be seen in Fig.~\ref{mla}(b). 
Note that considering the effect of  $ f (R) $ gravity the absorption is still increased in relation to the Schwarzschild black hole case. 
By comparing the amplitudes of the graphs of Fig. \ref{mla}, it is noted that the maximum amplitude of Fig. \ref{mla}(a) has a width narrower  than of Fig. \ref{mla}(b).

\begin{figure}[htbh]
\centering
\subfigure[]{\includegraphics[scale=0.93]{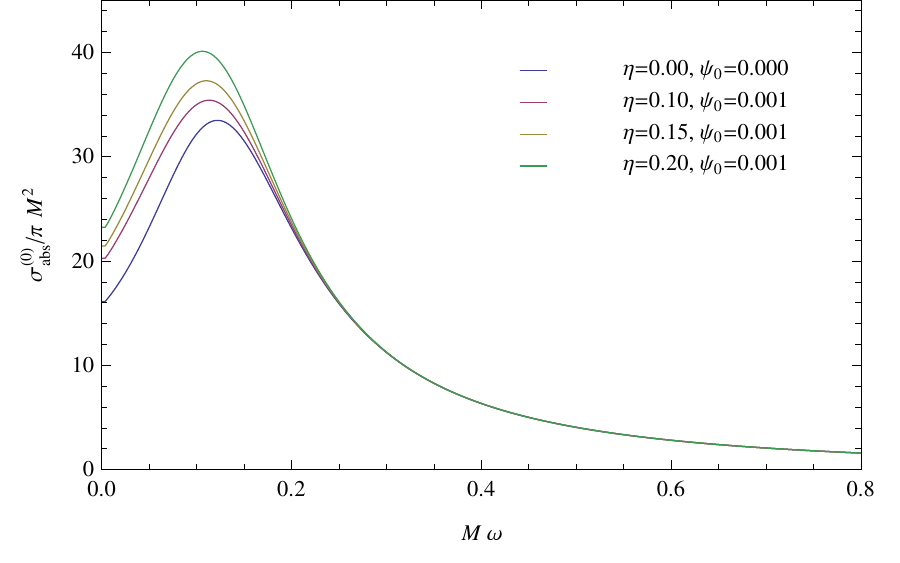}}\label{mlb1}
\qquad
\subfigure[]{\includegraphics[scale=0.93]{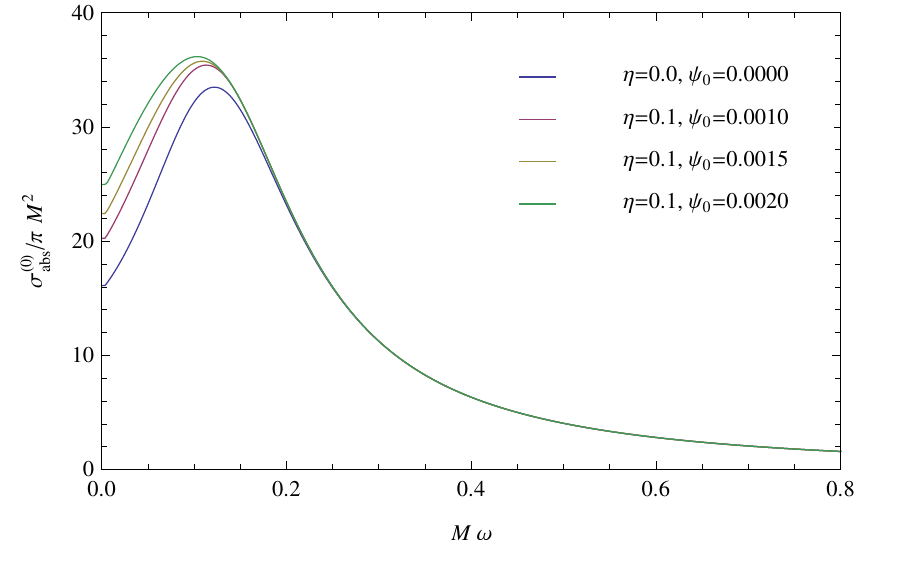}}\label{mlb2}
\caption{Partial absorption cross section for the $l=0$ mode. }
\label{mlb}
\end{figure}

Now considering the contributions of both the global monopole and the $f(R)$ gravity the graph \ref{mlb} shows a shift of the upward curve greater than in the previous curve.
\begin{figure}[htb]
\centering
{\includegraphics[scale=1.2]{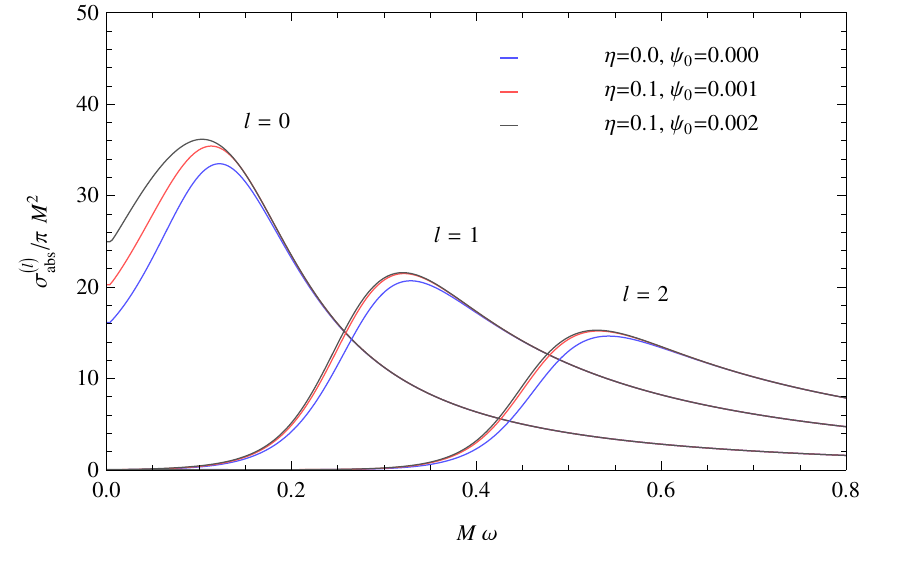}}
\caption{Partial absorption cross section to the modes $l=0,1,2$.}
\label{mlc}
\end{figure}
In Fig.~\ref{mlc} we plot the contribution of partial absorption to the modes $l = 0,1,2$. 
Note that for the modes $ l=1 $ and $ l=2 $ the partial absorption starts from zero and reaches a maximum value and then decreases with the increase of the energy $ M\omega $.
We can see that by increasing the value of  $l$ the corresponding maximum value of the partial absorption decreases. 
Therefore, our results are in accord with those obtained by the authors in~\cite{Hai:2013ara} and~\cite{LPing2015}, for instance. 
Furthermore, by analyzing the curves of Fig. \ref{mlc} we observe that as we increase the values of $ \psi_0 $  the amplitude is increased and this increase is greater for the $l = 0$ mode.

\section{Conclusions}
In summary,  in the present study we calculate the absorption and  scattering cross section of a black hole with a  global monopole in $f(R)$ gravity in the low-frequency limit at small angles ($ \theta\approx 0 $). 
To determine the phase shift analytically we have implemented the approximation formula $ \delta_{l}\approx (l-\ell) $ and so 
we have found, adopting the partial wave approach, that the scattering cross section is still dominated at the small-angled limit by $ 1/\theta^4 $. This dominant term is modified by the presence of the parameters $ \eta $ and $ \psi_0 $. 
Initially the case of a black hole with a global monopole was analyzed and we showed that the result for the differential scattering cross section as well as the absorption cross section is increased due to the monopole effect. {Moreover, considering the case of a black hole with a global monopole in $f(R)$ gravity, we find that in the low-frequency limit  the contribution to the dominant term  of the differential scattering cross section/absorption cross section is also increased due to the effect of the $ f(R)$ gravity. }
{Finally, we solve numerically the radial equation in order to calculate the partial absorption cross section for arbitrary frequencies. 
As a result we have shown that the absorption has its value increased as we increase the value of the parameter $\psi_0$.
}

\acknowledgements
We would like to thank CNPq and CAPES for partial financial support.


\begin{thebibliography}{100}
\bibitem{Frolov} A.V. Frolov, K.R. Kristjansson, L. Thorlacius et al, Phys. Rev. D {\bf 72}, 021501 (2005), 
[hep-th/0504073]; 

\bibitem{Townsend1997} P. K. Townsend, Black holes: Lecture notes, (University of Cambridge, Cambridge, 1997) [gr-qc/9707012]; T. Padmanabhan, Phys. Rep. {\bf 406}, 49 (2005), [gr-qc/0311036].

\bibitem{BezerradeMello:1996si} 
  E.~R.~Bezerra de Mello and C.~Furtado,
  Phys.\ Rev.\ D {\bf 56}, 1345 (1997).
  doi:10.1103/PhysRevD.56.1345

\bibitem{Yu2002} H. Yu, Phys. Rev. D {\bf 65}, 087502 (2002) .

\bibitem{Paulo2009}  J. Paulo, M. Pitelli and P. Letelier, Phys. Rev. D {\bf 80}, 104035 (2009) .

\bibitem{Chen2008} S. Chen and J. Jing, Mod. Phys. Lett. A {\bf 23}, 359 (2008).

\bibitem{Rahaman2005} F. Rahaman, P. Ghosh, M. Kalam and K. Gayen, Mod. Phys. Lett. A {\bf 20}, 1627 (2005).


\bibitem{Barriola1989} M. Barriola and A. Vilenkin, Phys. Rev. Lett. {\bf 63}, 341 (1989).

\bibitem{Kibble1976} T. W. B. Kibble, J. Phys. A {\bf 9}, 1387 (1976).

\bibitem{Vilenkin1988} A. Vilenkin, Phys. Rep. 121, 263 (1985).

\bibitem{Nojiri} S. Nojiri, S. D. Odintsov, Phys. Rev. D {\bf 68}, 123512 (2003) .

\bibitem{Carrol2004} S. M. Carrol, V. Duvvuri, M. Trodden, M. S. Turner, Phys. Rev. D {\bf 70}, 043528 (2004).

\bibitem{Fay2007}  S. Fay, R. Tavakol, S. Tsujikawa, Phys. Rev. D {\bf 75}, 063509 (2007). 

\bibitem{Bazeia:2007jj} 
  D.~Bazeia, B.~Carneiro da Cunha, R.~Menezes and A.~Y.~Petrov,
  Phys.\ Lett.\ B {\bf 649}, 445 (2007)
  doi:10.1016/j.physletb.2007.04.040
  [hep-th/0701106].

\bibitem{Carames2011} T. R. P. Carames, E. R. B. de Mello, M. E. X. Guimaraes, Int. J. Mod. Phys. Conf. Ser. 03, 446 (2011);
T. R. P. Carames, E. R. B. de Mello, M. E. X. Guimaraes, Mod. Phys. Lett. A {\bf 27}, 1250177 (2012).

\bibitem{Graca:2015jea} 
  J.~P.~Morais Gra\c ca, H.~S.~Vieira and V.~B.~Bezerra,
  Gen.\ Rel.\ Grav.\  {\bf 48}, no. 4, 38 (2016)
  doi:10.1007/s10714-016-2024-7
  [arXiv:1510.07184 [gr-qc]]; 
  J.~P.~Morais Graca and V.~B.~Bezerra,
  Mod.\ Phys.\ Lett.\ A {\bf 27}, 1250178 (2012).
  doi:10.1142/S0217732312501787;
  V.~B.~Bezerra and N.~R.~Khusnutdinov,
  Class.\ Quant.\ Grav.\  {\bf 19}, 3127 (2002)
  doi:10.1088/0264-9381/19/12/302
  [gr-qc/0204056].
  
\bibitem{Man2013} J. Man, H. Cheng, Phys. Rev. D {\bf 87}, 044002 (2013).

\bibitem{Lustosa:2015hwa} 
  F.~B.~Lustosa, M.~E.~X.~Guimarães, C.~N.~Ferreira and J.~L.~Neto,
  arXiv:1510.08176 [hep-th].

\bibitem{Man2015} J. Man, H. Cheng, Phys. Rev. D {\bf 92}, 024004 (2015).

\bibitem{Hai:2013ara} 
  H.~Hai, W.~Yong-Jiu and C.~Ju-Hua,
  Chin.\ Phys.\ B {\bf 22}, no. 7, 070401 (2013).

\bibitem{Futterman1988} J. A. Futterman, F. A. Handler, and R. A. Matzner,
{\it Scattering from black holes} (Cambridge University Press, England, 1988)
  
\bibitem{Matzner1977}  R. A. Matzner and M. P. Ryan, Phys. Rev. D {\bf 16}, 1636 (1977).

\bibitem {Westervelt1971} P. J. Westervelt, Phys. Rev. D {\bf 3}, 2319 (1971).

\bibitem{Peters1976} P. C. Peters, Phys. Rev. D {\bf 13}, 775 (1976).

\bibitem{Sanchez1976} N. G. S\'anchez, J. Math. Phys. {\bf 17}, 688 (1976);
N. G. S\'anchez, Phys. Rev. D {\bf 16} , 937 (1977);
N. G. S\'anchez, Phys. Rev. D {\bf 18}, 1030 (1978);
N. G. S\'anchez, Rev. D {\bf 18}, 1798 (1978).

\bibitem{Logi1977} W. K. de Logi and S. J. Kov\'acs, Phys. Rev. D {\bf 16}, 237 (1977).

\bibitem{Doram2002} C. J. L. Doran and A. N. Lasenby, Phys. Rev. D 66, 024006 (2002).

\bibitem{Dolan:2007ut} 
  S.~R.~Dolan,
  Phys.\ Rev.\ D {\bf 77}, 044004 (2008)
  doi:10.1103/PhysRevD.77.044004
  [arXiv:0710.4252 [gr-qc]].

\bibitem{Crispino:2009ki} 
  L.~C.~B.~Crispino, S.~R.~Dolan and E.~S.~Oliveira,
  Phys.\ Rev.\ D {\bf 79}, 064022 (2009)
  doi:10.1103/PhysRevD.79.064022
  [arXiv:0904.0999 [gr-qc]].


\bibitem{Churilov1974} A. A. Starobinsky and S. M. Churilov, Sov. Phys.- JETP {\bf 38}, 1 (1974).

\bibitem{Gibbons1975} G. W. Gibbons Commun. Math. Phys. {\bf 44}, 245 (1975)

\bibitem{Page1976} D. N. Page, Phys. Rev. D {\bf 13}, 198 (1976)

\bibitem{Unruh1976} W. G. Unruh, Phys. Rev. D {\bf 14}, 3251 (1976)

\bibitem{Churilov1973}  A. A. Starobinskii and S. M. Churilov, Zh. Eksp. Teor. Fiz. {\bf 65}, 3 (1973).

\bibitem{Crispino:2007zz} 
  L.~C.~B.~Crispino, E.~S.~Oliveira and G.~E.~A.~Matsas,
  Phys.\ Rev.\ D {\bf 76}, 107502 (2007).

\bibitem{Dolan:2009zza} 
  S.~R.~Dolan, E.~S.~Oliveira and L.~C.~B.~Crispino,
  Phys.\ Rev.\ D {\bf 79}, 064014 (2009)

\bibitem{Oliveira:2010zzb} 
  E.~S.~Oliveira, S.~R.~Dolan and L.~C.~B.~Crispino,
  Phys.\ Rev.\ D {\bf 81}, 124013 (2010).
 

 

\bibitem{Dolan}S. R. Dolan, E. S. Oliveira, L. C. B. Crispino, Phys. Lett. B {\bf 701}, 485 (2011).

\bibitem{ABP2012-1} M.~A.~Anacleto, F.~A.~Brito and E.~Passos, Phys. Rev. D {\bf 86}, 125015 (2012) 
[arXiv:1208.2615 [hep-th]]; Phys. Rev. D {\bf 87}, 125015 (2013) [arXiv:1210.7739 [hep-th]].

  
\bibitem{Anacleto:2015mta} 
  M.~A.~Anacleto, I.~G.~Salako, F.~A.~Brito and E.~Passos,
  Phys.\ Rev.\ D {\bf 92}, no. 12, 125010 (2015)
  doi:10.1103/PhysRevD.92.125010
  [arXiv:1506.03440 [hep-th]]; 
  M.~A.~Anacleto, F.~A.~Brito, A.~Mohammadi and E.~Passos,
  arXiv:1606.09231 [hep-th].  
  
  
\bibitem{Brito2015} M.~A.~Anacleto, F.~A.~Brito and E.~Passos,
  Phys.\ Lett.\ B {\bf 743}, 184 (2015)
  [arXiv:1408.4481 [hep-th]].
  
\bibitem{Jung2004} E. Jung and D. Park, Class. Quantum Grav. {\bf 21}, 3717  (2004), arXiv:hep-th/0403251 [hep-th];
E. Jung, S. Kim, and D. Park, Phys. Lett. B {\bf 602}, 105 (2004), arXiv:hep-th/0409145 [hep-th].  

\bibitem{Doran2005} C. Doran, A. Lasenby, S. Dolan, and I. Hinder, Phys. Rev. D {\bf 71}, 124020 (2005), 
arXiv:gr-qc/0503019 [gr-qc].

\bibitem {Dolanprd2006} S. Dolan, C. Doran, and A. Lasenby, Phys. Rev. D {\bf 74}, 064005 (2006), arXiv:gr-qc/0605031 [gr-qc].

\bibitem{Castineiras2007} J. Castineiras, L. C. Crispino, and D. P. M. Filho, Phys. Rev. D {\bf 75}, 024012 (2007).
 
\bibitem{Benone:2014qaa} 
  C.~L.~Benone, E.~S.~de Oliveira, S.~R.~Dolan and L.~C.~B.~Crispino,
  Phys.\ Rev.\ D {\bf 89}, no. 10, 104053 (2014)
  doi:10.1103/PhysRevD.89.104053
  [arXiv:1404.0687 [gr-qc]].
  
\bibitem{Moura:2011rr} 
  F.~Moura,
  JHEP {\bf 1309}, 038 (2013)
  doi:10.1007/JHEP09(2013)038
  [arXiv:1105.5074 [hep-th]].
  
  
\bibitem{Marinho:2016ixt} 
  C.~I.~S.~Marinho and E.~S.~de Oliveira,
  arXiv:1612.05604 [gr-qc].
  
  
  \bibitem{Vilenkin1985} A. Vilenkin, Phys. Rep. 121, 263 (1985).
  
\bibitem{Chen:2016ftz} 
  L.~Chen and H.~Cheng,
 Gen.\ Rel.\ Grav.\  {\bf 50}, no. 3, 26 (2018) arXiv:1607.07138 [hep-th]. 


\bibitem{Dolan:2012yc} 
  S.~R.~Dolan and E.~S.~Oliveira,
  Phys.\ Rev.\ D {\bf 87}, no. 12, 124038 (2013)
  [arXiv:1211.3751 [gr-qc]]. 




\bibitem {Yennie1954} D. R. Yennie, D. G. Ravenhall, and R. N. Wilson, Phys. Rev. 95, 500 (1954).

\bibitem{Cotaescu:2014jca} 
  I.~I.~Cotaescu, C.~Crucean and C.~A.~Sporea,
  Eur.\ Phys.\ J.\ C {\bf 76}, no. 3, 102 (2016)
  doi:10.1140/epjc/s10052-016-3936-9
  [arXiv:1409.7201 [gr-qc]].
  
\bibitem{Das:1996we} 
  S.~R.~Das, G.~W.~Gibbons and S.~D.~Mathur,
  Phys.\ Rev.\ Lett.\  {\bf 78}, 417 (1997)
  doi:10.1103/PhysRevLett.78.417
  [hep-th/9609052]. 
     
\bibitem{LPing2015} Liao Ping, Zhang Ruan-Jing, Chen Ju-Hua and Wang Yong-Jiu, Chin. Phys. Lett. {\bf 32}, No.5, 050401 (2015).


\end{thebibliography}
\end{document}